\documentclass[a4paper,twocolumn,superscriptaddress]{quantumarticle}
\pdfoutput=1
\usepackage[utf8]{inputenc}\usepackage[english]{babel}
\usepackage[T1]{fontenc}
\usepackage{amsmath}\usepackage{hyperref}
\usepackage{graphicx}\usepackage{amssymb}
\usepackage{physics}\usepackage{units} 
\usepackage{upgreek}\usepackage{tikz}\usepackage{lipsum}
\begin{document}
\title{Quantum probes for quantum wells}
\date{\today}
\author{Ilaria Pizio}
\email{ilaria.pizio@studenti.unimi.it}
\affiliation{Quantum Technology Lab, Dipartimento di Fisica {\it Aldo Pontremoli}, Universit\'a degli studi di Milano, I-20133 Milano, Italia}
\author{Shivani Singh}
\email{shivanis@imsc.res.in}
\affiliation{The Institute of Mathematical Sciences, C.I.T campus, Taramani, Chennai, 600113, India}
\affiliation{Homi Bhabha National Institute, Training School Complex, Anushakti Nagar, Mumbai 400094, India}
\author{C. M. Chandrashekar}
\email{chandru@imsc.res.in}
\affiliation{The Institute of Mathematical Sciences, C.I.T campus, Taramani, Chennai, 600113, India}
\affiliation{Homi Bhabha National Institute, Training School Complex, Anushakti Nagar, Mumbai 400094, India}
\author{Matteo G. A. Paris}
\email{matteo.paris@fisica.unimi.it}
\orcid{0000-0001-7523-7289}
\affiliation{Quantum Technology Lab, Dipartimento di Fisica {\it Aldo Pontremoli}, Universit\'a degli studi di Milano, I-20133 Milano, Italia}
\affiliation{The Institute of Mathematical Sciences, C.I.T campus, Taramani, Chennai, 600113, India}
\maketitle
\begin{abstract} 
We seek for the optimal strategy to infer the width $a$ of 
an infinite potential wells by performing measurements on the 
particle(s) contained in the well. In particular, we address 
quantum estimation theory as the proper framework to 
formulate the problem and find the optimal quantum measurement, 
as well as to evaluate the ultimate bounds to precision.
Our results show that in a static framework the best strategy
is to measure position on a delocalized particle, corresponding 
to a width-independent quantum signal-to-noise ratio (QSNR), 
which increases with delocalisation. Upon considering time-evolution 
inside the well, we find that QSNR increases as $t^2$. On the other 
hand, it decreases with $a$ and thus time-evolution is 
a metrological resource only when the width is not too large compared
to the available time evolution.
Finally, we consider entangled probes placed into the 
well and observe super-additivity of the QSNR: it is the 
sum of the single-particle QSNRs, plus a positive definite 
term, which depends on their preparation and may increase with
the number of entangled particles. 
Overall, entanglement represents a resource for the precise 
characterization of potential wells.
\end{abstract}
\section{INTRODUCTION}
In undergraduate Quantum Mechanics courses, the potential wells 
are usually the first examples used to illustrate quantum effects 
due to confinement and interference \cite{qw01,qw02,qw03,qw04,qw05,qw06}. 
On the other hand, quantum well (QW) potentials are not just 
an academic exercise. Rather, 
they are important models used in several branches of 
physics, since they often provide a surprisingly accurate 
description of different physical systems. At the same time, it 
may be used to illustrate potential drawbacks in canonical standard quantization \cite{cqw1,cqw2,cqw3}.
\par
In nuclear physics, where short range forces are dominant, QW 
potentials help to illustrate several phenomena at low energy 
\cite{qwn}. QWs 
are also employed to describe the confinement of electrons 
inside crystals, e.g., quantum wells, wires and dots 
corresponding to confinement in one, two or three dimensions, 
respectively \cite{qwqd1,qwqd2}. Those structures may be created 
by inserting in a given semiconductor a nano sized impurity made of a different one. 
Quantum dots, in particular, received much attention, because of 
their applications in nanoelectronics. 
The size of the quantum dot is a crucial parameter, since it 
determines the optical properties of the crystal; the smaller 
the dots, the larger is the intensity of the emitted light. 
As a consequence, the precise knowledge of the dimensions 
of the potential well, in particular of its width, is 
a crucial information for the development of effective light sources.
\par
In this paper, we consider a toy problem with potential applications
in the fields mentioned above. We consider an infinite QW in one 
dimension and seek for the optimal strategy to infer its width, 
denoted by $a$, by using {\it quantum probes} \cite{qp1,qp2,qp3,qp4,qp5,qp6,qp7,qp8,qp9,qp10,qp11}, i.e
performing measurement on the particles 
subjected to the QW potential. The analogue problem in $N$ 
dimensions may be then reduced to $N$ problems in one dimension. 
In particular, we address quantum estimation theory as the proper
framework where to formulate the problem and to find the optimal
quantum measurement, as well as to evaluate the ultimate quantum 
bounds to precision.
More precisely, we will consider one or more particles in a QW
and look for the optimal strategy to infer its width, i.e. we 
are looking for the best initial preparation, the optimal 
interaction time, and the more informative measurement, providing 
overall the highest precision in the determination of the width 
of the QW. In this optimization procedure, the figures of merit 
is the so-called quantum Fisher information, which provides a 
quantitative measure of the information about a parameter, which 
is extractable by any measurement performed on a family of 
quantum states.
\par
Our results show that in a static framework, position measurement
is the optimal one for any initial state, since its Fisher 
information is equal to the quantum Fisher information. In other 
words, position data provides us with all the available information 
about the width of well. Moreover, we found that before making a 
measurement it may be convenient to wait for a certain amount 
of time, because the quantum Fisher information increase with the time 
evolution  as $t^2$. Finally, we found that entanglement represents a resource, since precision may be enhanced using
multi-particle entangled probes.
\par
The paper is structured as follows. In Section 2, we briefly 
review the infinite square well quantum problem in one dimension
and provide an introduction to the ideas and the methods of 
quantum estimation theory, also evaluating the Fisher information 
for two relevant observables: position and energy.
In Section 3, we focus to static situations and evaluate 
the quantum Fisher information for different families of states, 
showing that delocalisation is the key feature to gain information
about the width of the well. In Section 4, we take into account 
time evolution and evaluate the quantum Fisher information for some
class of states. In Section 5, we address the use of  
$N$-particle probe to infer the width of the QW and 
describes how entangled probes may be used 
to improve precision at fixed number of particles. 
Section 6 closes the paper by summarising results 
and some concluding remarks.
\section{PRELIMINARY CONCEPTS}
\begin{figure}[h!]
\centering 
\includegraphics[width=.9\columnwidth]{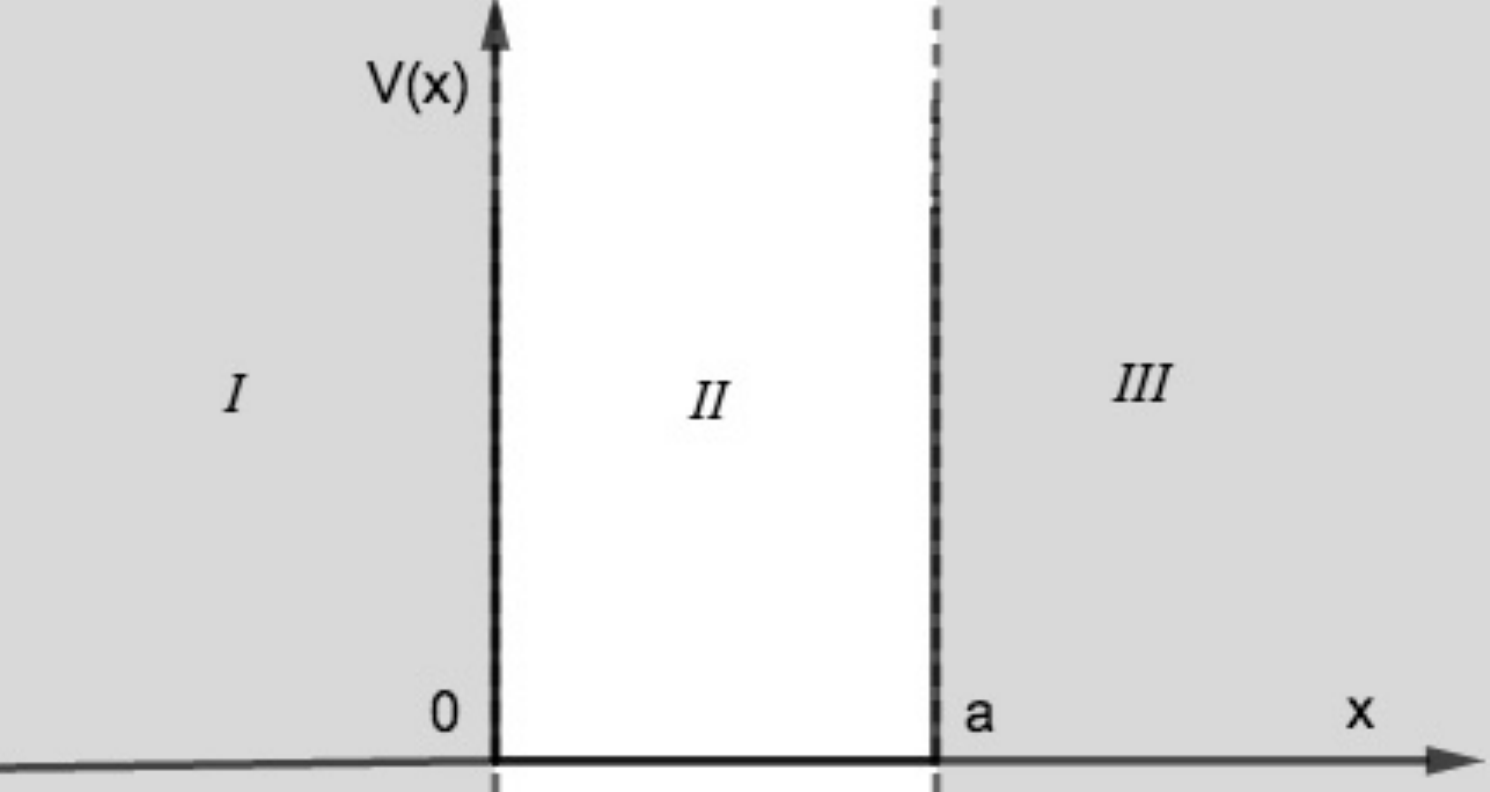}
\caption{The infinite square well potential, i.e. the 
model of a particle confined in the region between $x=0$ 
and $x=a$. Regions I and III are indeed forbidden 
because the potential is there infinite.} 
\label{potfig}
\end{figure}

In order to introduce the problem and establish notation, let 
us first review the infinite square well potential problem in
non-relativistic quantum mechanics, i.e. let us find the 
eigenvalues and the eigenfunctions of the one-dimensional 
Hamiltonian $H=\frac{p}{2m} + V(x)$, where the potential, 
of width $a$, is shown in Fig. \ref{potfig}, i.e. 
\begin{equation}
V(x)=
\begin{cases}
\infty & \text{ for } x < 0 \text{ and } x > a \\
0 & \text{ for } 0 \le x \le a .
\end{cases}
\label{potenziale}
\end{equation} 
We look for the eigenvalues and the eigenfunctions of the 
Hamiltonian $H$ by solving the correspondent eigenvalue 
equation in the position basis, i.e. solving the Schrodinger 
equation $- \frac12 \partial^2_x \psi(x) = [E-V(x)]\psi(x)$
for the wave-function $\psi(x)$, where we use natural unit 
$\hbar=1$, and assume unit mass $m=1$ for the particle.
The solution is straightforward upon dividing the space into
three regions (see Fig. \ref{potfig}) to see that regions 
I and III are forbidden because the potential there is 
infinite. The eigenfunctions form a discrete and 
non-degenerate spectrum of bound  
which may be written as
$$
|\psi_n\rangle = \int_0^a\!\! dx\, \psi_n(x)\,|x\rangle\,, \;
\int_0^a\!\! dx\, |\psi_n(x)|^2=1\,,
$$
where $n \in {\mathbb N}^+$ and 
\begin{align}
\label{autofunz}
\psi_n(x) & = \sqrt{2 \over a}\ \sin \left( 
\frac{n \pi}{a}x \right)\,,\\ 
E_n & = \frac{n^2 \pi^2 }{2  a^2}\,.
\label{autoval}
\end{align} 
The eigenfunctions exist only in the region II, i.e.  
all the integrals must be done between 0 and $a$, 
and form a complete orthonormal set. We remind that 
$n \in {\mathbb N}^+$, i.e. it cannot assume the 
value $n=0$, because in that case $\psi_0(x)=0$ and 
the uncertainty relations would be violated. The ground
state has energy $E_1 = \frac{\pi^2 \hbar^2}{2 ma^2}$, 
which also represents the energy splitting between the eigenstates.
\par
According to the Stone-Von Neumann theorem \cite{stvn}, 
the eigenstates of 
the Hamiltonian and the position eigenstates form two 
{\it unitarily inequivalent} basis. We will use both in 
the following of the paper, and write a generic state 
$|f\rangle$ as 
\begin{align}
|f\rangle & =  \int_0^a\!\! dx\, f(x) \ket{x}\,, \\
& = \sum_{n=1} ^\infty f_n\, |\psi_n\rangle\,,
\end{align}
where $f(x)=\langle x|f\rangle$, $f_n = \langle \psi_n|f\rangle$
and
\begin{align}
f(x) & = \sum_{n=1} ^\infty c_n \,\psi_n(x) \\ 
c_n & =  \int_0 ^a\!\! dx\, \psi_n(x)\, f(x)\,.
\end{align} 
Position basis, being independent on the value of the 
potential width, is suitable for the evaluation of 
the Fisher information and the signal-to-noise ratio (see the 
following Section), whereas the Hamiltonian basis is of course
the privileged one for time evolution. If we prepare a particle
in an initial state $|f_0\rangle=\sum_n  f_n\, |\psi_n\rangle$, 
the evolved state at time $t$ is given by 
\begin{align}
|f\rangle &= U_t |f_0\rangle \equiv \exp\{- i H t\} 
| f_0\rangle \notag \\ &= \sum_n e^{-i E_n t} f_n\, |\psi_n\rangle
\,. \label{ft}
\end{align}
\subsection{Quantum estimation theory}
It often happens in science that a quantity of interest 
is not accessible directly. Perhaps, the most prominent 
example in physics is that of a field, either gravitational, 
magnetic, or electric. As a matter of fact, no device is 
actually measuring, e.g., the magnetic field. Rather, one 
measures the effect of the field on a moving charge, say 
measuring its acceleration, deflection or displacement, 
and then {\em estimate} the field by suitably processing 
the data observed for the measured quantity.
\par
The chosen measurement and the data processing are together 
referred to as the {\em inference strategy} for the parameter 
of interest $\xi$ \cite{b1,b2,b3,b4}. 
After a certain observable $X$ has been 
chosen, the available data ${\mathbf x}=(x_1,x_2,.....,x_M)$ 
is a set of outcomes from $M$ repeated measurements of $X$, 
i.e. a sample from the distribution $p({\mathbf x}|\xi) = 
\Pi_{k=1}^M p(x_k|\xi)$, which itself depends on the 
parameter that has to be estimated. The estimated value 
for $\xi$ is the average value of an estimator
\begin{equation}
\bar \xi = \int\!\! d{\mathbf x}\, p({\mathbf x}|\xi)\,
\hat \xi ({\mathbf x})\,,
\end{equation}
i.e. a map $\hat \xi \equiv \hat \xi ({\mathbf x})$
from the space of observations to the space of the 
parameters. The overall precision of the estimation 
procedure is quantified by the variance of $\hat\xi$, i.e.
\begin{equation}
\hbox{Var}\, \hat \xi = \int\!\! d{\mathbf x}\, p({\mathbf x}|\xi)\,
[\xi({\mathbf x})-\bar \xi]^2\,.
\end{equation}
The variance of any unbiased estimator 
(i.e. an estimator for which $\bar \xi \rightarrow T$ in 
the asymptotic limit $M \gg1$) for the parameter $T$ 
is bounded by Cramer-Rao theorem \cite{cr1,cr2,cr3}, 
stating that
\begin{equation}
\hbox{Var}\,\hat\xi \ge \frac{1}{ M F(\xi)}
\end{equation}
where $F(\xi)$ is the Fisher information (FI)
\begin{equation}
F(\xi)= \int\! dx \ p(x|\xi) \left[ \frac{\partial
\log p(x|\xi) }{\partial \xi} \right]^2 \,,
\label{fi}
\end{equation}
$p(x|\xi)$ being the single outcome probability, i.e. 
the probability of measuring $x$ when the true value 
of the parameter is $\xi$. The FI quantifies the amount 
of information about the parameter $\xi$ that we may 
extract from the measurement of $X$. 
\par
In a quantum mechanical setting, the conditional probability 
$p(x|\xi)$ is given by the Born rule $p(x|\xi)= \Tr[P_x \rho_{\xi}]$, 
where $\rho_{\xi}$ the density operator describing the (parameter-dependent) state of the system and $P_x$ is the projector over 
the eigenstate of a selfadjoint operator $X$ corresponding 
to the eigenvalue $x$.
\par
In order to write the Fisher information in a convenient form, 
and to maximise its value over the possible observables, we
introduce the Symmetric Logarithmic Derivative (SLD) $L_{\xi}$, 
i.e. a selfadjoint operator satisfying the equation 
\begin{equation}
\frac{L_{\xi} \rho_{\xi} + \rho_{\xi} L_{\xi}}{2} = \frac{\partial \rho_{\xi}}{\partial \xi}\,. \label{sld}
\end{equation}
Upon inserting Eq. (\ref{sld}) in Eq. (\ref{fi}) we may find 
an upper bound for the FI of any quantum measurement
\begin{equation}
F(\xi) \le \Tr[\rho_{\xi}\, L_{\xi}^2] = H(\xi)\,,
\end{equation}
which is usually referred to as the Quantum Fisher 
information (QFI) \cite{he68,bc94,bh98}, 
and coincides with the least monotone quantum 
Riemannian metric \cite{pe96}. 
An optimal estimation strategy should
employs  measurement with $F(\xi)=H(\xi)$ and then use 
an optimal estimator which saturates the quantum Cramer-Rao 
bound $\hbox{Var}\,\hat\xi \ge 1/ M H(\xi)$. An 
optimal measurement with $F(\xi)=H(\xi)$ is provided by
SLD itself \cite{ak06}, though other problem-specific measurements may
achieve similar precision.
\par
The precision of a parameter estimation strategy depends on the 
variance of the estimator. In order to compare different strategies,
we have also to consider the variance in terms of the mean value,
i.e. we should consider the signal-to-noise ratio (SNR) 
\begin{equation}
R_{\xi} = \frac{\xi^2}{\hbox{Var}(\xi)}\,
\end{equation}
that is larger for good strategies. The Cramer-Rao inequality 
bounds this quantity with the quantum signal-to-noise ratio 
(QSNR), defined as follows
\begin{equation}
R_{\xi} \le Q(\xi)=\xi^2 H(\xi)\,.
\end{equation}
The larger is $Q(\xi)$ the more {\it estimable} is in principle 
the parameter.
Overall, quantum estimation theory says that in order to
optimally estimate a parameter, we should find a state 
preparation with the largest QSNR and then measure the 
SLD, or any other observable with a FI as close as possible 
to the QFI. When the information about the parameter is 
encoded onto pure states $\rho_{\xi}=\ket{\psi_{\xi}}
\bra{\psi_{\xi}}$, one has  $\rho_{\xi}^2=\rho_{\xi}$ and 
the SLD may be easily found as 
\begin{equation}
L_{\xi} = 2\, \partial_{\xi} \rho_{\xi} = 2\, \Big[ \ket{\psi_{\xi}}\bra{\partial_{\xi} \psi_{\xi}} + \ket{\partial_{\xi} \psi_{\xi}}\bra{\psi_{\xi}} \Big]\,.
\end{equation}
The corresponding QFI is given by $H(\xi) = \bra{\psi_{\xi}}L_{\xi}^2\ket{\psi_{\xi}}$, i.e.  \cite{lqe}
\begin{align}
H(\xi) & = 4 \Big[
\bra{\partial_{\xi} \psi_{\xi}}\ket{\partial_{\xi} \psi_{\xi}} 
+ \left|\bra{\partial_{\xi} \psi_{\xi}}\ket{\psi_{\xi}}\right|^2 
\notag \\ 
& \quad \quad  +
\bra{\partial_{\xi} \psi_{\xi}}\ket{\psi_{\xi}}^2  
+ \bra{\psi_{\xi}}\ket{\partial_{\xi} \psi_{\xi}}^2 
 \Big] \,.
\label{qfi_puro}
\end{align}
\subsection{Single-particle quantum probes}
Using results from the previous Section, we now evaluate the 
information about $a$ contained into the state of a particle
placed into the well. In other words, we evaluate the
QFI of Eq. (\ref{qfi_puro}) for a generic pure state at time 
$t$, as in Eq. (\ref{ft}). In order to simplify notation, we 
will use the following shorthands
\begin{align}
\frac{\partial}{\partial a} \rightarrow \partial\,,
\quad 
\int_0^a\!\! dx \rightarrow \int \!\! dx\,, \quad
\sum_{n=1} ^\infty \rightarrow \sum_n\,.
\end{align}
At first, we need the state derivative with respect to 
the parameter $a$
\begin{align}
\ket{\partial f} &= \int \! dx  \sum _n \partial \left[f_n (x) e^{-i E_n t} \psi_n(x) \right] \ket{x}\end{align}
where $g(x,a,t) \equiv \partial \left[f_n\, e^{- i E_nt} 
\psi_n(x) \right]$ is  given by 
\begin{align}
g(x,a,t) = \psi_n \partial f_n + f_n \partial \psi_n  - i  t f_n 
\psi_n \partial E_n \,. \label{gg}
\end{align}
In Eq. (\ref{gg}) we have removed the explicit dependence on $a$, $x$ and $t$, whereas the derivatives of the eigenvalues and eigenfunctions are
given by 
\begin{align}
\partial E_n & = - \frac{n^2 \pi^2}{a^3}  \label{eiv}\\
\partial \psi_n & = - \frac{1}{2}\ \sqrt{\frac{2}{a^3}} \left[ 
\sin \frac{n \pi x}{a}  +  \frac{2 n \pi x}{a} \cos \frac{n \pi x}{a} \right]\,.
\end{align}
In order to proceed we need few scalar products. The first is
just the orthonormality of the Hamiltonian basis 
$\bra{\psi_m}\ket{\psi_n} =
\int\! dx\, \psi_n(x)\, \psi_m^*(x) = \delta_{mn}$ and the others
are 
\begin{align}
\label{eig_2}
\bra{\psi_m}\ket{\partial \psi_n} =& \frac2a\, (1-\delta_{mn})\, (-1)^{m+n}\frac{ m n}{n^2-m^2} \\
\bra{\partial \psi_m}\ket{\partial \psi_n}  =& 
(1-\delta_{mn})
\frac{(-1)^{m+n}}{a^2} \frac{4\, n\, m\, (m^2 + n^2)}{(m^2-n^2)^2}  \notag \\ 
&+ \delta_{mn}\frac{1}{a^2} \left(\frac{n^2 \pi^2}{3} + 
\frac{1}{4}\right) \,.
\label{eig_3}
\end{align}
We also notice that $\bra{\psi_m}\ket{\partial \psi_n}$ is 
anti-symmetric for the exchange of $n$ and $m$ whereas $\bra{\partial \psi_m}\ket{\partial\psi_n}$ is symmetric. Using this symmetry it
is easy to prove that 
$\bra{f}\ket{\partial f}$ is a purely imaginary quantity at any time and
for any choice of the initial state, whereas
$\bra{\partial f}\ket{\partial f}$ is a real quantity.
Overall, we have that the QFI in Eq. (\ref{qfi_puro}) may 
be rewritten as
\begin{equation}
H(a)= 4 \left[ \bra{\partial f}\ket{\partial f}+ 
\bra{f}\ket{\partial f}^2 \right]\,,
\label{qfi_real}
\end{equation}
where the second term is real and negative.
\subsection{FI for some relevant measures}
Let us focus on the static case, i.e. we assume that a particle
is placed in the well, prepared in a given quantum state $\ket{f}$,
and then an observable is immediately measured, without leaving
the particle {\it to move} within the well, i.e. its quantum state
to evolve. We also assume without loss of generality (see below) that
the wave function $f(x)= \bra{x}\ket{f}$ is real.
In this conditions $\bra{\partial f}\ket{f}=0$ and $H(a) = 
4 \int\! dx\, (\partial f)^2$. On the other hand, the probability distribution in a position measurement is given by $p(x|a) = |f(x)|^2$ 
and thus its Fisher information is 
\begin{align}
F(a) &= \int \! dx \; \frac{1}{|f(x)|^2} \left[\partial |f(x)|^2\right]^2 \\ 
&= 4 \int \! dx \, \left[ \partial f(x) \right]^2 \,,
\end{align}
which is equal to the QFI for any choice of (real) $f(x)$. If $f(x)$ is complex the line of reasoning is the same, though a rotation should be made to the state before measuring position.
\par
Another relevant measurement is that of energy. 
The possible outcome are the eigenvalues of the Hamiltonian, 
and the probability distribution is given by
\begin{equation}
p(E_n|a) = |\bra{\psi_n}\ket{f}|^2 = |f_n|^2.
\end{equation}
The FI for the energy measurement is thus given by
\begin{align}
F(a) &= \sum_n  \frac{1}{|\langle\psi_n |f\rangle|^2} \Big[ 
\partial|\bra{\psi_n}\ket{f}|^2 \Big]^2 \notag \\
& = 4 \, \sum_n  \Big[ \partial  |f_n| \Big]^2\,.
\label{fi_energia}
\end{align}
The energy FI is not, in general, equal to the QFI. In particular, 
it is useless to prepare the particle in an eigenstate of the 
Hamiltonian, since the only possible result is the correspondent eigenvalue with probability 1 and the measure does not give any information about the parameter $a$. On a generic state, we gain
some information from an energy measurement if the expansion 
coefficients does depend on the parameter $a$.
\section{STATIC PROBES}
Let us start our analysis with different possible single-particle preparations and by focussing on the static case, i.e. we assume that
a particle is placed in the well in a given quantum state $\ket{f}$,
and that an observable is immediately measured, without leaving
the particle the time to evolve. 
\par
We start considering an eigenstate of the Hamiltonian and 
calculate the QFI and the QSNR. Since the eigenfunctions are 
reals, $\bra{f}\ket{\partial_a f} = 0$ and the Q-quantities 
reads as follows
\begin{align}
\label{h_eigen}
H_n(a) &= 4 \bra{\partial \psi_n}\ket{\partial \psi_n}  
= \frac{3 + 4 n^2 \pi^2}{3 a^2} \\
\label{q_eigen}
Q_n  &= 1+ \frac{4}{3} n^2 \pi^2 = 1 + \frac83 a^2 E_n.
\end{align}
The QSNR does not depend on $a$ and increases with $n$, which means
that in principle we should prepare the particle in an eigenstate 
with large $n$ in order to gain information about the parameter $a$.
\par
We next consider a superposition of two generic eigenstates
\begin{align}
\ket{f_{nm}} = \cos \alpha \ket{\psi_n} + \sin \alpha \ket{\psi_m}\,.
\label{sovrapposizione}
\end{align}
Upon straightforward calculations we have 
\begin{align}
Q_{nm}(\alpha) = &  \cos^2\alpha\, Q_n +\sin^2\alpha\,  
Q_m \notag \\  & + a^2 \sin 2\alpha\, \langle\partial \psi_n|
\partial \psi_m\rangle\,,
\end{align}
where $\langle\partial \psi_n|
\partial \psi_m\rangle$ is given in Eq. (\ref{eig_3}). Also in this
case the QSNR does not depend on $a$. We also notice that 
$Q_{nm} (\alpha) = Q_{mn} (\pi/2-\alpha)$ and thus in the following
we consider $m = n+d > n$ and $0\leq \alpha \leq \pi/4$.
\par
In order to properly assess the effects of superpositions 
we fix the overall energy of the state and compare the 
QSNR of $\ket{f_{nm}}$ with $Q_{[[\bar n]]}$, where 
$E_{\bar n}$ is the mean energy of the superposition state 
$\ket{f_{nm}}$, i.e. $E_{\bar n} = E_n \cos^2\alpha + E_{n+d} 
\sin^2\alpha$, and $[[x]]$ denotes the round function, i.e. the 
closest integer to $x$. We have $n \leq \bar n \leq n+d$ where
\begin{align}
\bar n = \sqrt{n^2 \cos^2\alpha + (n+d)^2 \sin^2\alpha}\,.
\end{align}
A remarkable result may be obtained by considering 
unbalanced superposition corresponding to small values of 
$\alpha$. In this case, upon defining 
$\gamma_{nd} (\alpha) = Q_{n,n+d} (\alpha)/Q_{[[\bar n]]}$,
we have 
\begin{align}
\bar n & \stackrel{\alpha \ll 1}{\simeq} n + O(\alpha^2) \label{sp1}\\
\gamma_{nd} (\alpha) & \stackrel{\alpha 
\ll 1}{\simeq} 1+ (-1)^d\, g_{nd}\,\alpha + O(\alpha^2) \label{sp2}\\
g_{nd} & = \frac{24 n (n+d) (d^2+2nd+2n^2)}{d^2 (d+2n)^2(3+4 n^2 \pi^2)} >0\notag\,.
\end{align}
Eqs. (\ref{sp1}) and (\ref{sp2}) says that with a negligible 
increase of energy, and preparing the particle in a superposition 
with even $d$, one may increase the QSNR by a non-negligible amount.
Notice that at fixed $n$, $g_{nd}$ decreases with $d$, and thus the
most convenient superposition is the state $|f_{n,n+2}\rangle$. 
\begin{figure}[h!]
\centering
\includegraphics[width=.9\columnwidth]{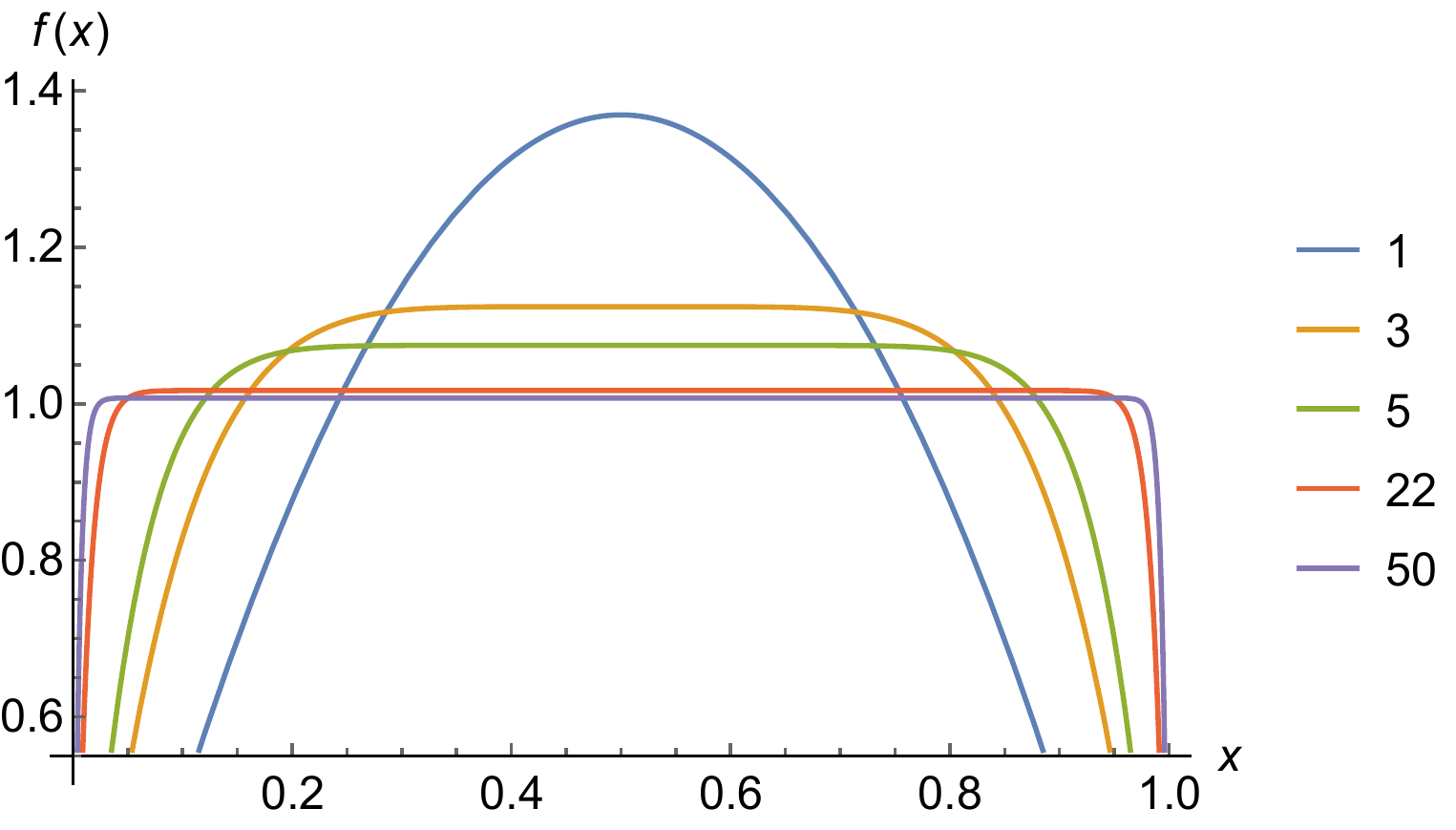}
\caption{The wave function in Eq. (\ref{poli}) for different 
values of the parameter $p$. The function becomes very flat 
already for small values of $p$.}
\label{polinomio}
\end{figure}
\par
The examples above suggest that the delocalisation of the particle
inside the well may play a role in increasing the QSNR. This agrees
with intuitive arguments based on the fact that position measurement 
is optimal, and thus the more delocalised is the particle, the 
more information may be gained from a position measurement. In order to
make this reasoning more quantitative, let us consider the 
family of states $|f_p\rangle$ where the wave function is given by 
\begin{equation}
f_p(x;a)= N[-(2x-a)^{2p} + a^{2p}]
\label{poli}
\end{equation}  
where $p \in {\mathbb N}^+$, $p>1$, and 
the normalization factor is given by
\begin{equation}
N = \sqrt{\frac{1 +6p + 8p^2}{8p^2 a^{1+4p}}}\,.
\end{equation} 
The wavefunction in Eq. (\ref{poli}) becomes more and more 
flat for increasing $p$, approaching a box function for 
large $p$. In Fig. \ref{polinomio}, we show the behaviour 
of $f_p(x;a)$ for different values of $p$ and for $a=1$. 
Upon exploiting the scaling $f_p(x;a)= 1/\sqrt{a} f_p(x/a;1)$ 
the behaviour for a generic value of the width may be recovered.
Concerning the QSNR, after straightforward calculations we have
\begin{align}
Q_p &= \frac{(1+4p)(1+8p)}{(4p-1)}\,,
\label{q_polino}
\end{align}
which is independent on $a$ and it is an increasing
function of $p$. We have $Q_1=15$ and $Q_p \simeq 8p$ for 
large $p$ ($p \gtrsim 10$ is already enough). 
\par
\begin{figure}[h!]
\includegraphics[width=.9\columnwidth]{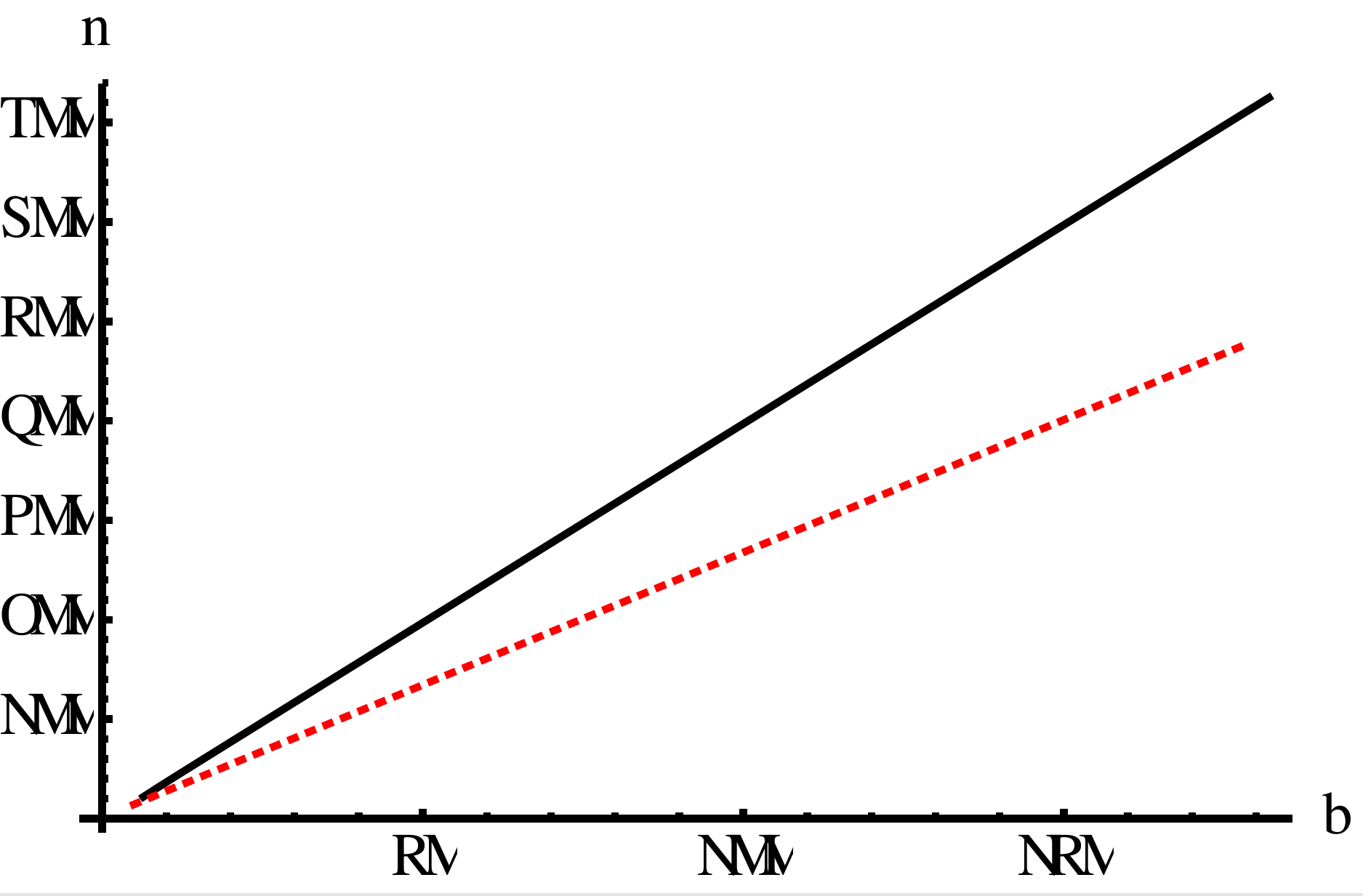}
\caption{The QSRN as a function of energy (expressed in unit of $1/a^2$) 
for Hamiltonian eigenstates (red dotted line) and for the delocalised 
states $|f_p\rangle$ (black solid line).}
\label{cdloc}
\end{figure}

 
The average energy of a $p$-state is given by
\begin{align}\label{ep}
\langle f_p|H|f_p\rangle = \frac1{a^2} \frac{1+6p+8 p^2}{4p-1}\,,
\end{align}
and thus, using Eqs. (\ref{ep}) and (\ref{eiv}), we have that 
$|f_p\rangle$ and $|\psi_n\rangle$ have the same energy if 
$$
\frac{1+6p+8 p^2}{4p-1} = \frac{n^2 \pi^2}{2}\,.
$$
In turn, this means that at fixed energy, the delocalised states 
$|f_p\rangle$ provide more information than the 
Hamiltonian eigenstates. This is illustrated in Fig. \ref{cdloc}, 
where we show the QSRN as a function of energy (in unit of $1/a^2$) 
for both families of states. 
\par
Similar conclusions may be obtained by considering some specific 
measure of delocalisation, e.g. the differential entropy of the
position distribution $p(x) = |f(x)|^2$ for different classes of 
states.
\section{DYNAMICAL PROBES}
In practice, it is not possible to prepare a system and perform
a measurement instantaneously. As a consequence, a question arises 
on how the information about the width of the well changes with time. 
In this Section, we will introduce time evolution and 
analyze whether this degree of freedom may be exploited to increase
the QFI. Intuitively, one may expect evolution to be beneficial, 
since, roughly speaking, the wavefunction does not have the possibility 
to get out of the well and thus should interact with the walls of the
well many times, accumulating more information about the structure 
of the potential.
\par
At first, let us check whether for eigenstate of the Hamiltonian information is left unchanged. If we prepare the particle
in an eigenstate $|\psi_n\rangle$ at time $t$ the state of the 
system is given by 
\begin{align}
\psi_n(x,t) &= \sqrt{\frac{2}{a}} \sin \left( \frac{n \pi}{a}x \right) e^{- i E_n t } \,,
\end{align}
so that 
\begin{align}
\ket{\partial_a \psi_n(t)}= e^{- i E_n t } \Big[ \ket{\partial_a \psi_n } -\,i\, t\, (\partial_a E_n) \ket{\psi_n} \Big]\,,
\end{align}
(notice that in this case, the wave function is not real anymore
and thus, in general, $\bra{\psi}\ket{\partial\psi} \neq 0$).
The QFI is given 
\begin{align}
H(a) 
&= 4 \Big[ \bra{\partial \psi_n(t)}\ket{\partial  \psi_n(t)} +  \bra{\psi_n(t)}\ket{\partial  \psi_n(t)}^2 \Big] \\
 & = \frac{3+4n^2\pi^2}{3a^2}\,,
\end{align}
which is indeed unchanged, compared to the static case. 
\par
\begin{figure}[h!]
\centering 
\includegraphics[width=.9\columnwidth]{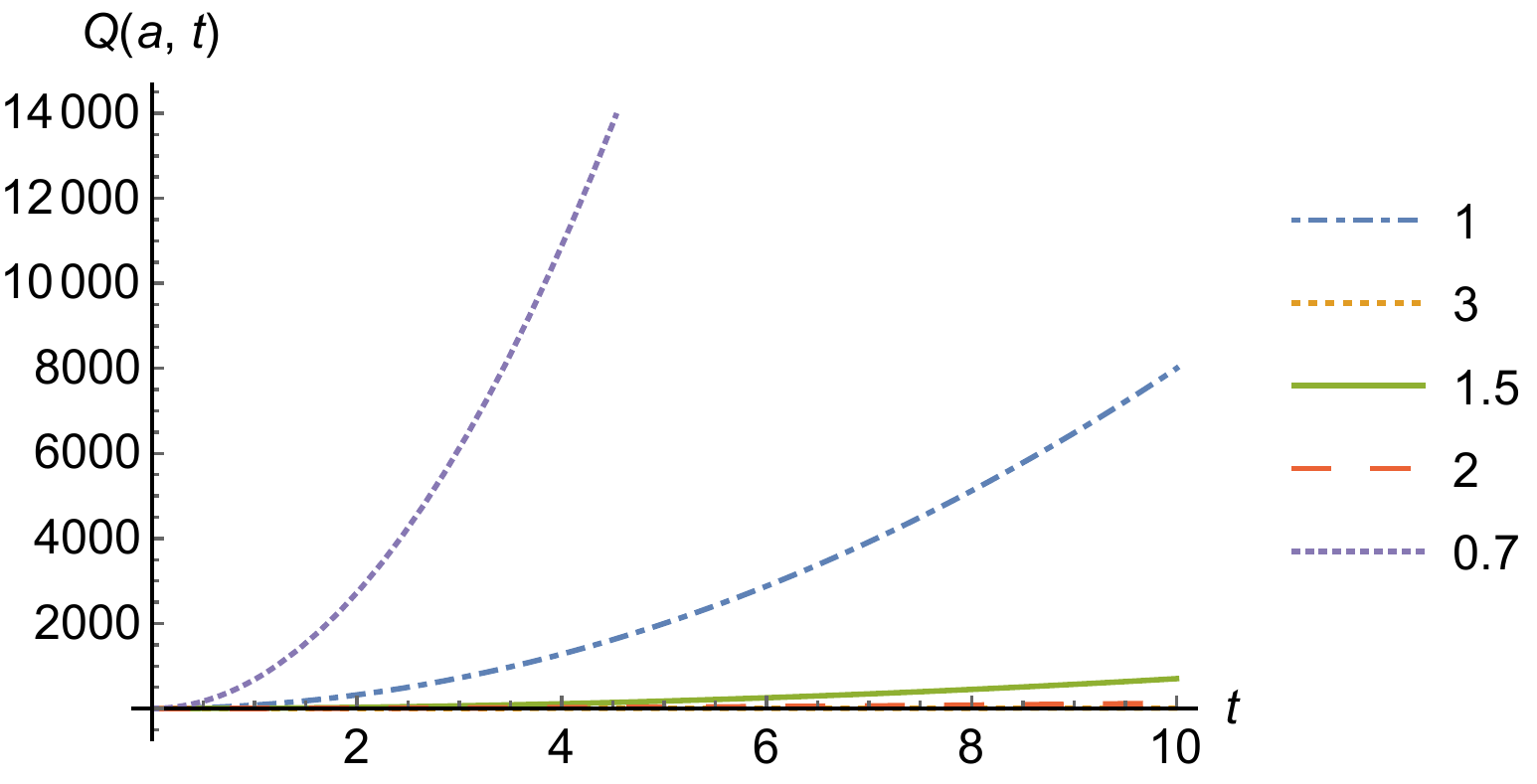}
\caption{The QSTN $Q(a,t)=a^2 H(a,t)$ see Eq.(\ref{h_par_tempo}) as function of time for different values of the parameter $a$. Numerical results are obtained by truncating the sum in Eq.(\ref{h_par_tempo}) at $n,m=50$, corresponding to a residual error $\epsilon \lesssim 10^{-6}$. The plot shows that the QSTN increase quadratically with time an decrease with the width $a$.} 
\label{q_parab_tempo}
\end{figure}

Let's now consider a generic initial preparation, which 
evolves as
\begin{align}
\ket{f(t)} = \sum_n f_n \ket{\psi_n} e^{- i E_n t }  \quad 
f_n = \bra{\psi_n}\ket{f} \in {\mathbb R}\,.
\end{align}
We do not report the full expression of the QFI and
rather assume that the amplitudes $f_n$ do not depends on $a$, 
i.e. are determined by external operations. In this case the
QFI rewrites as 
\begin{widetext}
\begin{align}
H(a,t) = & 4 \Bigg\{ t^2\ \sum_n  f_n^2\, (\partial E_n)^2  - 
\left(   \sum_n t\, f_n ^2\, \partial_a E_n   + \sum_{n m}   \sin(\Delta_{nm}t) \, f_n f_m  \langle\psi_m|\partial\psi_n\rangle    \right)^2 \notag
\\ & +     \sum_{n m}  \cos(\Delta_{nm}t) f_n f_m \langle\partial\psi_m|\partial\psi_n\rangle  + f_n f_m t \sin(\Delta_{nm} t)\, (\partial E_m + \partial E_n) 
\langle\psi_m|\partial\psi_n\rangle
            \Bigg\}\,, \label{longH}
\end{align}
where $\Delta_{nm} = E_n-E_m$. In turn, Eq. (\ref{longH}) 
suggests a $t^2$ dependence of the QFI. 
\par
In order to see these feature in a quantitative way, let us consider a simple initial state $|f\rangle$ with wave function of the form $f(x) = \sqrt{30/a^5} x (a-x)$, corresponding to amplitudes $f_n=0$ when $n$ 
is even and $f_n=8 \sqrt{15}/n^3 \pi^3$ if $n$ is odd.
Inserting this expression in the QFI of Eq. (\ref{longH}) we have
\begin{align}
H(a,t) &= 4\, \Bigg\{120\,\frac{t^2}{a^6}  + 1920   \sum_{n m} 
(-1)^{m+n} \frac{(m^2+n^2)}{a^2 \pi^4\, n^2\, m^2}
\left[    \frac{2\,\cos(\Delta_{nm}t)}{\pi^2\, (m^2-n^2)^2}  -  
\frac{t\,\sin(\Delta_{nm}t) }{a^2 (m^2-n^2)}  \right]  \notag \\ 
&- \left(-10\,\frac{t}{a^3} + 1920 \sum_{n m}\, (-1)^{n+m}   \frac{\sin(\Delta_{nm}t) }{a\, n^2\, m^2\, \pi^6 (m^2-n^2)}   
\right)^2   
\Bigg\} \label{h_par_tempo}
\end{align}
where all the sums include odd values only. 
\end{widetext}
Upon expanding the QSNR $Q(a,t)=a^2 H(a,t)$ for short times,
one obtains the leading term $Q(a,t) \sim t^2/a^4$, showing 
that when the state particle evolves within the well the QSNR
increases as $t^2$. On the other hand, a dependence on the width
itself appears, making dynamical probes convenient if the well is
not too large, compared to the available interaction time.
In Fig. \ref{q_parab_tempo} we show the QSNR $Q(a,t)$ as a function
of time for different values of the width $a$. As it is apparent
from plot, the expansion $Q(a,t) \sim t^2/a^4$ well describes the behaviour of $Q(a,t)$ also when the interaction time is not so small.
Results are obtained by numerically performing the sums of Eq.(\ref{h_par_tempo}) up to $n,m=50$, corresponding to a residual 
error $\epsilon \lesssim 10^{-6}$.
\par
The same behaviour may be observed with different preparation of
the particle. Overall, we found that, in general, the amount of 
information about the parameter $a$ increases quadratically in 
time at any fixed value of $a$. At the same time, 
evolution brings a dependence on the width itself, making more
and more difficult to estimate its value as it increases.
\section{ENTANGLED PROBES}
In the previous Sections, we have considered a single particle as
in the well as a quantum probe for its width. In this Section we 
address the use of more than one particle, and, in particular, of 
$N$ particles prepared in an entangled state. For the sake of simplicity, 
we take the particles identical, but distinguishable, and non interacting. 
We will start from two-particle probes, and then generalise the analysis
to $N$ particles.
\subsection{Two-particle entangled probes}
In the case of two particles, the  total Hamiltonian is given 
by $H_{tot}=H_1+H_2+V$, where 
$H_i$ the kinetic term $p_i ^2 /2m_i$ and the potential is that 
of Eq. (\ref{potenziale}). The eigenstates are 
the tensor products $\ket{\psi_i} \otimes \ket{\psi_j}$ 
of the single-particle eigenstates of Eq.(\ref{autofunz}), and 
the eigenvalues $E_{tot}$ are the sum $E_i+E_j$ of the eigenvalues 
in Eq.(\ref{autoval}).
\par
For any two-particle state 
\begin{align}
|f\rangle\rangle = \int\!\!\int\! dx_1 dx_2\, f(x_1, x_2)\, |x_1\rangle\otimes |x_2\rangle\,,
\end{align} 
and assuming a real wave-function $f(x_1, x_2)$ the QFI is given by 
$H(a)=4 \int\!\!\int\! dx_1 dx_2\, [\partial f(x_1,x_2)]^2$, thus confirming that also for two particles the (joint) measurement of
position is an optimal measurement. Notice that the measurement of
the position of only one of the particles is not optimal.
\par
Let us now consider the two particles prepared in an 
energy-particle entangled state, i.e. in a superposition 
state where we do not know which particle is in which 
(Hamiltonian) eigenstate. The wave-function is given by
\begin{align}
\Psi(x_1,x_2) = \frac{\psi_{n_1}(x_1)\psi_{n_2}(x_2) + 
\psi_{n_1}(x_2)\psi_{n_2}(x_1)}{\sqrt{2}}\,.
\label{entangled}
\end{align}
Following the procedure outlined in the previous Sections, 
we may easily evaluate the QSNR, which may be expressed as 
\begin{equation}\label{qen}
Q_{n_1n_2}^{(2)}= Q_{n_1}+Q_{n_2} + 32 \, 
\frac{n_1^2 \, n_2^2}{(n_1^2-n_2^2)^2}\,,
\end{equation}
where $Q_{n_1}$ and $Q_{n_2}$ are the single particle QSNRs given 
in Eq. (\ref{q_eigen}). Eq. (\ref{qen}) contains a remarkable result: 
the QSNR obtained using two particles in an entangled state is always
greater that the QSNR obtained using the two particles in two successive
experiments. 
\par
Motivated by the results of Section 3, let us now consider  
two-particle probes prepared in an entangled state of two
single-particle delocalised states of Eq.(\ref{poli}) 
with different indices, i.e.
\begin{align}
f_{p_1p_2}(x_1,x_2) = &  \frac{1}{\sqrt{2}} \big[f_{p_1}(x_1;a)f_{p_2}(x_2;a) \notag \\ & +f_{p_1}(x_2;a)f_{p_2}(x_1;a) \big]\,.
\end{align}
The corresponding QSNR is given by
\begin{align}
Q^{(2)}_{p_1p_2} = &\,   Q_{p_1} + Q_{p_2}  \\
& + \frac{(1+4p_1)(1+4p_2)(1+4p_1+4p_2)}{2(4p_1^2 +4p_2^2+8 p_1p_2-1)}\,,
\notag
\end{align}
where $Q_{p_1}$ and $Q_{p_2}$ are given in Eq.(\ref{q_polino}). 
The additional term is positive definite also in this case, i.e. 
entanglement leads to superadditivity of the QFI and the QSNR. 
In order to evaluate quantitatively the improvement, let us introduce
the ratio
\begin{align}
\gamma_{p_1p_2}=\frac{Q^{(2)}_{p_1p_2}}{Q_{p_1} + Q_{p_2}} > 1\,.
\end{align}
\begin{figure}[h!]
\centering
\includegraphics[width=.9\columnwidth]{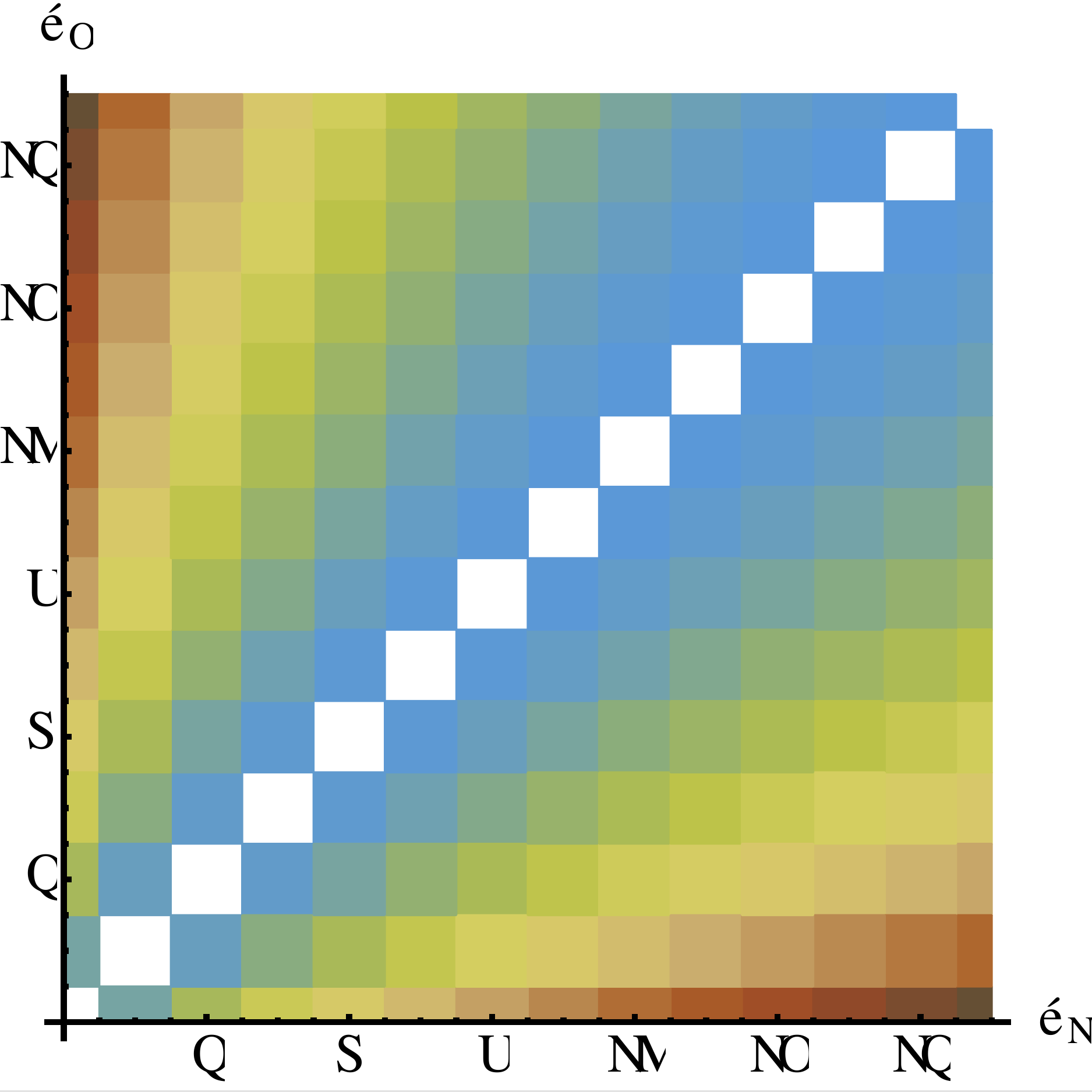}
\caption{The ratio $\gamma_{p_1p_2}$ as a function of $p_1$ 
and $p_2$ in the range $p=2,..,15$ ($\gamma_{pp}$ is undefined). 
The plot tells us that the entanglement makes the QSNR super-additive 
and the QSNR of a two-particle entangled states is always larger 
than the sum of the two single-particle QSNRs.}
\label{mat}
\end{figure}
\par
In Fig. \ref{mat} we show $\gamma_{p_1p_2}$ as a function of $p_1$ and $p_2$ in the range $p=2,..,15$ ($\gamma_{pp}$ is undefined) 
and notice that it achieves its maximum value $\gamma \simeq 5/4$ 
for $p_2 = p_1 \pm 1$. For increasing values of both the indices the
region in which $\gamma$ is close to its maximum becomes larger and larger.
\subsection{N-particle entangled probes}
Given the results of the previous Section, a question 
arises on whether using more particles one may achieve 
a better precision. The answer is positive, as it may easily be
shown upon considering the following three-particle entangled
probe, prepared in a $W$-like state with a wave-function 
of the form
\begin{align}\label{entw3}
\Psi(x_1,x_2,x_3) =  \frac1{\sqrt{3}} & \Big[
\psi_{n_1}(x_1)\psi_{n_1}(x_2)\psi_{n_2}(x_3) 
\\ & + 
\psi_{n_1}(x_1)\psi_{n_2}(x_2)\psi_{n_1}(x_3) 
\notag \\ & + 
\psi_{n_2}(x_1)\psi_{n_1}(x_2)\psi_{n_1}(x_3) 
\Big]\notag\,.
\end{align}
Upon exploiting Eq. (\ref{eig_2}) one arrives at 
\begin{align}
H^{(3)}_{n_1n_2} = \frac13 \Big[\, 6 H_{n_1} (a) + &\, 
3 H_{n_2} (a) \\ 
+& 48\, \big| \langle\partial\psi_{n_1} 
| \psi_{n_2}\rangle \big|^2\Big]\,,\notag
\end{align}
where $H_n(a)$ is given in Eq. (\ref{h_eigen}).
The QSNR is thus given by
\begin{equation}\label{qenW3}
Q_{n_1n_2}^{(3)}= 2 Q_{n_1}+Q_{n_2} + 64 \, 
\frac{n_1^2 \, n_2^2}{(n_1^2-n_2^2)^2}\,,
\end{equation}
where we have the sum of the single-particle QSNRs and an additional positive definite term, which is twice the 
one obtained with two particles, see Eq. (\ref{qen}). 
In order to compare the two results in the high-energy 
regime $n_1,n_2 \gg 1$, let us consider the most 
convenient choice for both, i.e. $n_1 \rightarrow n$, $n_2 
\rightarrow 1+n$. In this case,
we have
\begin{align}
\frac{Q_{n,1+n}^{(3)}}{2 Q_{n}+Q_{1+n}} & \stackrel{n \gg 1}{=} 1 + \frac4{\pi^2} + O(\frac1{E_n})\\
\frac{Q_{n,1+n}^{(2)}}{Q_{n}+Q_{1+n}} & \stackrel{n \gg 1}{
=} 1 + \frac3{\pi^2}+ O(\frac1{E_n})\,.
\end{align}
The argument may be then generalized to more particles, thus confirming that entanglement is a resource in the estimation of the width, and that the enhancement may increase with 
the number of entangled particles.
\par
Notice, however, that precision strongly depends on the 
preparation of the probe, and that entanglement alone
is not enough to improve precision. In order to show
this explicitly, let us consider the case of 
$N$ distinguishable particles prepared 
in a GHZ-like state, i.e. with a wave-function given by 
\begin{align}\label{eqPsi}
\Psi(x_1,x_2,...) =& \frac{1}{\sqrt{2}} \Big[ \prod_{i = 1}^{N} \psi_{n_i}(x_i) + \prod_{j = 1}^{N} \psi_{m_j}(x_j) \Big]
\end{align}
where state $\mathbf{m}=\{m_1,...\}$ is a permutation of $\mathbf{n}=
\{n_1,...\}$ and $\psi_k(x)$ is the $k$-th eigenstate of the Hamiltonian.
The wave-function is real and so $H(a) = 4\bra{\partial_a \Psi}\ket{\partial_a \Psi}= 4 (I_1 + I_2)/a^2$ where 
\begin{align}
I_1 =  a^2 & \int \prod_{i=1}^N dx_i\, \Bigg\{ \Big[ \sum_{j=1}^N \partial \psi_{n_j}(x_j) \prod_{l \neq j} \psi_{n_l}(x_l) \Big] \notag \\
\times & \Big[ \sum_{k=1}^N \partial \psi_{n_k}(x_k) \prod_{h \neq k} \psi_{n_h}(x_h) \Big] \Bigg\}\,, \label{eqI1} \\
I_2 = a^2 & \int \prod_{i=1}^N dx_i\, \Bigg\{ \Big[ \sum_{j=1}^N \partial \psi_{n_j}(x_j) \prod_{l \neq j} \psi_{n_l}(x_l) \Big] \notag \\
\times & \Big[ \sum_{k=1}^N \partial \psi_{m_k}(x_k) \prod_{h \neq k} \psi_{m_h}(x_h) \Big] \Bigg\} \label{eqI2} \,.
\end{align}
Using results from previous Sections and after calculations,
we have
\begin{align} \label{eqRI1}
I_1 & = \sum_{j=1}^N \Big( \frac{n_{j}^2 \pi^2}{3} + \frac{1}{4} \Big)\,, \\ 
\label{eqRI2}
I_2 &= 4 \sum_{k,j=1}^{N}\! \delta_{m_k n_j}\delta_{m_j n_k}   \Big( \frac{n_k n_j}{n_{k}^2 - n_{j}^2} \Big)^2 \!\!\prod_{\hbox{\scriptsize$
\begin{array}{c} l \neq k \\ l \neq j\end{array}$}}\!\! \delta_{n_l m_l}\,. 
\end{align}
The corresponding QSNR is given by
\begin{align}\label{ghzq}
& Q^{\hbox{\tiny{(N)}}}_{n_1,n_2,...,n_N} = \sum_{j=1}^N Q_{n_j} + \\ &+ 16 \sum_{k,j=1}^{N}\! \delta_{m_k n_j}\delta_{m_j n_k}   \Big( \frac{n_k n_j}{n_{k}^2 - n_{j}^2} \Big)^2 \!\!\prod_{\hbox{\scriptsize$
\begin{array}{c} l \neq k \\ l \neq j\end{array}$}}\!\!  \delta_{n_l m_l}\,. \notag
\end{align}
As it is apparent from Eq. (\ref{ghzq}), the presence of "conflicting deltas" in the expression of $Q^{\hbox{\tiny{(N)}}}_{n_1,n_2,...,n_N}$
make it impossible to surpass the two-particle QSNR 
$Q_{n_1n_2}^{(2)}$ of Eq. (\ref{qen}) using 
$N$-particle GHZ-like states.
\section{CONCLUSIONS}
In this paper, we have used quantum estimation theory
as the proper framework to address the precise 
characterization of an infinite potential wells, 
i.e. the estimation of its width.
In particular, we have been looking for the optimal 
measurement to be performed on the particles in the 
well, and for their best preparation, in order to obtain 
the ultimate bound to precision, as imposed by 
quantum mechanics.
\par
In doing this we have evaluated the quantum Fisher 
information of the corresponding quantum statistical
models, and the Fisher information for selected kind 
of measurements. We have also considered different 
preparations of the system in order to illustrate the 
different features of the problem. Finally, we have 
evaluated the quantum signal-to-noise ratio (QSNR) in 
order to compare the different working regimes.
\par
Our results show that the best measurement we may 
perform on a static system, is the position measure, 
because in that case the FI equals the QFI for any 
state and any value of the width.
On the contrary, performing an energy measurement 
is useless unless one is able to prepare suitable superpositions with parameter dependent coefficients.
\par
In a static setting, the QSNR is independent
of the width, and the best way to initialise the system 
is to prepare it in a delocalized state, which could 
be an eigenstate of the Hamiltonian with a large 
eigenvalue, or a wave-function as the polynomial 
in Eq.(\ref{poli}).  We have then considered 
time evolution inside the wells and found that the QSNR increases with time as $t^2$. Letting the system evolve is thus convenient, since the amount of information  increases. 
On the other hand, the QSNR decreases with $a$ itself, 
and so time evolution is a resource only if the well is 
large enough compared to the available interaction 
time.
\par
Finally, we have considered $N$-particle probes and found 
that entanglement enhances precision, since the QSNR is 
the sum of the single-particle QSNRs plus a positive definite 
term, which depends on state preparation, and may increase 
with the number of entangled particles.
\section*{Acknowledgements}
This work has been supported by SERB through project 
VJR/2017/000011. MGAP is member of GNFM-INdAM.

\begin{thebibliography}{99}

\bibitem{qw01} M. Belloni, R.W. Robinett, {\it The infinite well and Dirac delta function potentials as pedagogical, mathematical and physical models in quantum mechanics}, Phys. Rep. {\bf 540}, 25 (2014).
\bibitem{qw02} I. Marzoli {\it et al.}, {\it 
Quantum Carpets made simple}, Acta Phys. Slov. {\bf 48}, 323 (1998).
\bibitem{qw03} C. U. Segre, J. D. Sullivan, {\it Bound-state wave packets}, Am. J. Phys. {\bf 44}, 729 (1976).
\bibitem{qw04} L. M. Ar\'evalo Aguilar {\it et al.}, {\it The infinite square well potential and the evolution operator method for the purpose of overcoming misconceptions in quantum mechanics}, Eur. J. Phys. {\bf 35} 025001 (2014).
\bibitem{qw05} F. Gori {\it et al.}, {\it The general wavefunction for a particle under uniform force}, Eur. J. Phys. {\bf 22}, 53 (2001).
\bibitem{qw06} S. Waldenstrom {\it et al.}, {\it The Force Exerted by the Walls of an Infinite Square Well on a Wave Packet: Ehrenfest Theorem, Revivals and Fractional Revivals}, Phys. Scr. {\bf 68}, 45 (2003).
\bibitem{cqw1} G. Bonneau {\it et al.}, {\it 
Self-adjoint extensions of operators and the teaching 
of quantum mechanics}, Am. J. Phys. {\bf 69}, 322 (2001).
\bibitem{cqw2} M. Znojil, {\it 
PT-symmetric square well}, Phys. Lett. A {\bf 285}, 7 (2001).
\bibitem{cqw3} P. L. Garcia de Leon {\it et al.}, {\it 
Infinite quantum well: A coherent state approach}, 
Phys. Lett. A {\bf 372}, 3597 (2008).
\bibitem{qwn} D. L. Hill and J. A. Wheeler, {\it Nuclear Constitution and the Interpretation of Fission Phenomena}, 
Phys. Rev. {\bf 89}, 1102 (1953).
\bibitem{qwqd1} Y. Alhassid, {\it The statistical theory of quantum dots}, Rev. Mod. Phys. {\bf 72}, 895 (2000).
\bibitem{qwqd2} M. Lozada-Cassou  {\it et al.}, {\it Quantum features of semiconductor quantum dots}, 
Phys. Lett. A {\bf 331}, 45 (2004).
\bibitem{qp1} A. Smirne {\it et al.}, {\it Quantum probes to assess correlations in a composite system}, 
Phys. Rev. A {\bf 88}, 012108 (2013).
\bibitem{qp2} C. Benedetti {\it et al.}, {\it Quantum probes for the spectral properties of a classical
environment}, Phys. Rev. A {\bf 89}, 032114 (2014).
\bibitem{qp3} M. G. A. Paris, {\it Quantum probes for fractional Gaussian processes}, Physica A {\bf 413}, 256 (2014).
\bibitem{qp4} C. Benedetti, M. G. A. Paris, {\it Characterization of classical Gaussian processes using quantum probes}, 
Phys. Lett. A {\bf 378}, 2495 (2014).
\bibitem{qp5} M. A. C. Rossi, M. G. A. Paris,  {\it Entangled quantum probes for dynamical environmental noise},  
Phys. Rev. A {\bf 92}, 010302(R) (2015).
\bibitem{qp6} D. Tamascelli {\it et al.}, {\em Characterization of qubit chains by Feynman probes}, Phys. Rev. A {\bf 94}, 042129 (2016).
\bibitem{qp7} L. Seveso, M. G. A. Paris, {\em Can quantum probes satisfy the weak equivalence principle?}, 
Ann. Phys. {\bf 380}, 213 (2017).
\bibitem{qp8} M. Bina {\it et al.}, {\it Continuous-variable quantum probes for structured environments}, 
Phys. Rev. A \textbf{97} 012125 (2018)
\bibitem{qp9} C. Benedetti {\it et al.}, {\it 
Quantum probes for the cutoff frequency of Ohmic environments}, 
Phys. Rev. A \textbf{97} 012126 (2018)
\bibitem{qp10} F.  Troiani, M. G. A. Paris, {\it 
Universal quantum magnetometry with spin states at equilibrium}, 
Phys. Rev. Lett. \textbf{120}, 260503 (2018).
\bibitem{qp11}
A. Beggi {\it et al.}, {\it Probing the sign of Hubbard interaction by two-particle quantum walks}, Phys. Rev. 
A {\bf 97}, 013610 (2018).
\bibitem{stvn} B. C. Hall, {\it The Stone-von Neumann 
Theorem} in {\it Quantum Theory for Mathematicians}, 
Grad. Texts Math. {\bf 267}, (Springer, New York, 2013), p. 279.
\bibitem{b1} G. Casella and R. L. Berger, {\it Statistical inference}, 
Vol. 2 (Duxbury Pacific Grove, CA, 2002).
\bibitem{b2} H. L. Van Trees and K. L. Bell, {\it Detection estimation and modulation theory}, pt. I (Wiley, 2013).
\bibitem{b3} S. M. Kay, {\it Fundamentals of statistical signal processing} (Prentice Hall PTR, 1993).
\bibitem{b4} E. L. Lehmann and G. Casella, {\it Theory of point estimation} (Springer, Berlin, 2006).
\bibitem{cr1} H. Cram\`er, {\it Mathematical Methods of Statistics} (PMS-9), Vol. 9 (Princeton University Press, 2016).
\bibitem{cr2} C. R. Rao, in {\it Breakthroughs in statistics} 
(Springer, Berlin, 1992) pp. 235?247.
\bibitem{cr3} R. A. Fisher, in {\it Mathematical Proceedings of the Cambridge Philosophical Society}, Vol. 22 
(Cambridge University Press, 1925) p. 700.
\bibitem{he68} C. Helstrom, {\it The minimum variance of estimates in quantum signal detection}, 
IEEE Trans. Inf. Theory {\bf 14}, 234 (1968).
\bibitem{bc94} S. L. Braunstein and C. M. Caves, {\it Statistical distance and the geometry of quantum states}, 
Phys. Rev. Lett. {\bf 72}, 3439 (1994).
\bibitem{bh98} D. C. Brody, L. P. Hughston, {\it Statistical geometry in quantum mechanics}, Proc. Roy. Soc. A {\bf 454}, 2445 (1998). 
\bibitem{pe96} D. Petz, {\it Monotone metrics on matrix spaces}, Lin. Alg. Appl. {\bf 244}, 81 (1996).
\bibitem{ak06} A. Fujiwara, {\it Strong consistency and asymptotic efficiency for adaptive quantum estimation 
problems}, J. Phys. A {\bf 39}, 12489 (2006).
\bibitem{lqe} M. G. A. Paris, {\it Quantum estimation for quantum technology}, Int. J. Quantum Inf. {\bf 07}, 125 (2009).


\end{thebibliography}
\end{document}